# Discrete-time quantum walks on one-dimensional lattices

Xin-Ping Xu

*School of Physical Science and Technology,*
*Suzhou University, Suzhou, Jiangsu 215006, P.R. China*

(Dated: March 9, 2010)

## Abstract

In this paper, we study discrete-time quantum walks on one-dimensional lattices. We find that the coherent dynamics depends on the initial states and coin parameters. For infinite size of lattice, we derive an explicit expression for the return probability, which shows scaling behavior $P(0,t) \sim t^{-1}$ and does not depends on the initial states of the walk. In the long-time limit, the probability distribution shows various patterns, depending on the initial states, coin parameters and the lattice size. The average mixing time $M_\epsilon$ closes to the limiting probability in linear $N$ (size of the lattice) for large values of thresholds $\epsilon$. Finally, we introduce another kind of quantum walk on infinite or even-numbered size of lattices, and show that the walk is equivalent to the traditional quantum walk with symmetrical initial state and coin parameter.

PACS numbers: 03.67.Lx, 05.40.Fb, 03.65.Db, 03.67.Ca

## I. INTRODUCTION

Random walk is related to the diffusion models and is a fundamental topic in discussions of Markov processes. Several properties of (classical) random walks, including dispersal distributions, first-passage times and encounter rates, have been extensively studied. The theory of random walk has been applied to computer science, physics, ecology, economics, and a number of other fields as a fundamental model for random processes in time [1–3].

Quantum random walk, which is a natural extension of the classical random walk, has attracted a great deal of attention in the scientific community in recent years. The continuous interest in the study of quantum random walk can be partly attributed to its broad applications in the field of quantum information and computation [4, 5]. Quantum random walks can be used to design highly efficient quantum algorithms for quantum computer [6, 7]. For example, Grovers algorithm can be combined with quantum walks in a quantum algorithm for glued trees which provides even an exponential speed up over classical methods [7–9]. Besides their important applications in quantum computation, quantum walks are also used to model the coherent exciton transport in solid state physics [10]. This could be done in the framework of the tight-binding model, which is equivalent to the so-called continuous-time quantum walk on discrete structures [11, 12]. It is shown that the dramatic non-classical behavior of quantum walks can be attributed to quantum coherence, which does not exist in the classical random walks.

In the literature, there are two types of quantum random walks: discrete-time and continuous-time quantum walks [13]. The main difference between them is that discrete time quantum walk requires a "coin"-which is just any unitary matrix-plus an extra Hilbert space on which the coin acts, while continuous-time quantum walks do not need this extra Hilbert space. Aside from this, these two versions are similar to their classical counterparts. Discrete-time quantum walks evolve by the application of a unitary evolution operator at discrete time intervals, and continuous-time quantum walks evolve under a (usually time-independent) Hamiltonian. Unlike the classical case, the extra Hilbert space for discrete-time quantum walks means that one cannot obtain the continuous quantum walk from the discrete walk by taking a limit as the time step goes to zero. This is because discrete time quantum walks need an extra Hilbert space, called the "coin" space, and taking the limit where the time step goes to zero does not eliminate this Hilbert space. Although there is no natural limit to go from the discrete to continuous quantum walks for general graphs, Ref. [14] offers a treatment of this limit for the quantum walk on the line, where it is possible to meaningfully extract the continuous-time walk as a limit of the discrete-time walk. The dynamics of quantum walks of both types has been studied in detail for walks on an infinite line-for the continuous-time case in Refs. [11, 15, 16] and for the discrete-time case in Refs. [17–20], it has been shown that the properties of discrete and continuous time quantum walks are different.

Here, we focus on discrete-time quantum walks (DTQWs). Previous work have studied DTQWs on the line and cycles. The behavior of DTQWs on the line is strikingly different from the classical random walks because of quantum interference. The variance $\sigma^2$ of the quantum walk is known to grow quadratically with the number of steps, $t$, $\sigma^2 \propto t^2$, compared to the linear growth, $\sigma^2 \propto t$, for the classical random walk [13, 19]. Since the cycle (or one-dimensional lattice) is a line segment with periodic boundary conditions, the solutions of quantum walks on cycles can be simplified greatly on consideration of the Fourier space of the particle [21]. For a classical random walk on a 1D lattice of size $N$, the mixing time $M_\epsilon$ converges to the uniform distribution in time $O(-N^2 ln\epsilon)$ [21]. Quantum mechanically, the probability oscillates forever and in general do not mix even instantaneously. However, by defining a time-averaged probability distribution, the quantum walks can mix to the uniform or non-uniform distribution. In Ref. [22], Bednarska et al. have studied the long-time limiting probability distribution for a Hadamard walk on 1D lattice (cycles). They have shown that the Hadamard walk converges to the uniform distribution on odd-numbered size of cycles but it converges to a nontrivial distribution on even-numbered 1D lattices. Previous studies related to DTQWs on 1D lattices (cycles) focus on a particular choice of the initial state and coin parameter [23–25]. Here, we consider DTQWs on 1D lattices for various initial states and coin parameters, and discuss the effect of these parameters on the properties of the quantum dynamics.

In this paper, we give a systematic study of DTQWs on 1D lattice with various initial states and coin parameters. We explore the time evolution of the walk, return probability, long-time limiting probability and mixing times, and compare the properties for various initial states and coin parameters. The paper is organized as follows: In Sec. II we briefly review discrete-time quantum walks on regular graphs. In Sec. III, we derive analytical results for DTQWs on 1D lattice and find an explicit formula for the return probability. We also do computer simulation to implement DTQWs on 1D lattice for various parameters, and find that the numerical results accurately agree with our analytical results. In Sec. IV, we study the properties of mixing times and discuss the influence of the parameters of the walk. In Sec. V, we define another kind of quantum walk on infinite or even-numbered size of lattices, and prove that the defined walk is equivalent to the traditional quantum walk with symmetrical initial state and coin parameter. Conclusions are given in the last part, Sec. VI.

## II. DISCRETE-TIME QUANTUM WALKS

Discrete-time quantum walk was first introduced by Mayer and Aharonov et al. in Ref. [26, 27]. Discrete-time quantum walk takes place in a discrete space of positions, with a unitary evolution of coin toss and position shift in discrete time steps. Here, we define discrete-time quantum walks (DTQWs) on $d$-regular graph, which is a regular graph each vertex has exactly $d$ edges.

The discrete-time quantum walk on $d$-regular graph happens on the coin Hilbert space $\mathcal{H}_c$ and position Hilbert space $\mathcal{H}_p$, the total Hilbert space is given by $\mathcal{H} = \mathcal{H}_c \otimes \mathcal{H}_p$ [13]. If the $d$-regular graph has $N$ vertices we have $\mathcal{H}_p = \{|i\rangle : i = 1, 2, ..., N\}$, $\mathcal{H}_c = \{|e_i\rangle : i = 1, 2, ..., d\}$. The coin flip operator $\hat{C}$ and position shift operator $\hat{S}$ are applied to the total state in $\mathcal{H}$ at each time step [13]. The coin flip operation $\hat{C}$ (only acting on $\mathcal{H}_c$) is the quantum equivalent of randomly choosing which way the particle will move, then the position-shift operation $\hat{S}$ moves the particle according to the coin state, transferring the particle into the new superposition state in position space. For every vertex, all the outgoing edges is labeled as $1, 2, \ldots, d$. The conditional shift operation $\hat{S}$ moves the particle from $v$ to $w$ if the edge $(v, w)$ is labeled by $j$ on $v$'s side [13]:

$$\hat{S}|e_j\rangle \otimes |v\rangle = \begin{cases} |e_j\rangle \otimes |w\rangle, & if \;\; e_v^j = (v, w) \\ 0, & otherwise. \end{cases} \quad (1)$$

The evolution of the system at each step of the walk is governed by the total operator,

$$\hat{U} = \hat{S}(\hat{I} \otimes \hat{C}), \quad (2)$$

where $\hat{I}$ is the identity operator in $\mathcal{H}_p$. Thus the total state after $t$ steps is given by,

$$|\psi(t)\rangle = \hat{U}^t|\psi(0)\rangle. \tag{3}$$

Finally, we obtain the probability distribution,

$$P(x,t) = \sum_{i=1}^{d} |\langle e_i, x|\psi(t)\rangle|^2 = \sum_{i=1}^{d} |\langle x| \otimes \langle e_i|\psi(t)\rangle|^2. \tag{4}$$

For $d$-regular graphs, the coin flip matrix $\hat{C}$ can be of various forms. The most common coins analyzed in the field are the Grover coin and the discrete Fourier transform (DFT) coin [13, 28]. It is shown that the choice of coin flip $\hat{C}$ and initial coin state strongly influence the behavior of discrete-time quantum walks. In the next section, we will concentrate DTQWs on 1D lattice. We will choose a general form of the initial coin state and coin flip matrix, and derive analytical results for the walk.

## III. DISCRETE-TIME QUANTUM WALKS ON ONE-DIMENSIONAL LATTICE

DTQWs on the line has already been analyzed in detail and the equivalence of all unbiased coin operators has been noted by several authors [17–20, 29–31]. Here, we continue to use this framework and extend the calculations for DTQWs on 1D lattices.

### A. Analytical solutions

In the following, we restrict our attention to DTQWs on 1D lattice. Without loss of generality, we consider the one-parameter family of coins,

$$\hat{C} = \begin{pmatrix} \sqrt{\rho} & \sqrt{1-\rho} \\ \sqrt{1-\rho} & -\sqrt{\rho} \end{pmatrix}, \quad 0 \leqslant \rho \leqslant 1 \tag{5}$$

The value of $\rho = 1/2$ corresponds to the Hadamard coins, which is a balanced coin and involves the coin into each direction in $\mathcal{H}_c$ with equal probability. Suppose the particle was initially ($t = 0$) localized at vertex $x_0$ and the initial coin states distributed in the coin subspace,

$$|\psi(0)\rangle = (\sqrt{a}|e_1\rangle + \sqrt{1-a}e^{i\phi}|e_2\rangle) \otimes |x_0\rangle, \tag{6}$$

where the two parameters $a \in [0,1]$ and $\phi \in [0, 2\pi)$. The position shift operator $\hat{S}$ has the following form [13],

$$\hat{S} = (\sum_i |i-1\rangle\langle i|) \otimes |e_1\rangle\langle e_1| + (\sum_i |i+1\rangle\langle i|) \otimes |e_2\rangle\langle e_2|. \tag{7}$$

The total states $|\psi(t)\rangle$ and probability distribution $P(x,t)$ after $t$ steps are determined by Eqs. (3) and (4). The solution of the problem can be greatly simplified in the Fourier space. The Fourier transformation of the state in particle Hilbert space can be written as,

$$|\widetilde{\psi}_N(k,t)\rangle = \frac{1}{\sqrt{N}} \sum_{x=1}^{N} e^{2\pi i k x/N}|\psi_N(x,t)\rangle, \quad x \in \{1, 2, ..., N\}. \tag{8}$$

Then the time evolution of the states in the Fourier picture turn into a single difference equation,
$$|\widetilde{\psi}_N(k,t)\rangle = \widetilde{U}(k)|\widetilde{\psi}_N(k,t-1)\rangle, \tag{9}$$
where $\widetilde{U}(k)$ is time evolution operator in the Fourier space,
$$\widetilde{U}(k) = \begin{pmatrix} \sqrt{\rho}e^{-2\pi ki/N} & \sqrt{1-\rho}e^{-2\pi ki/N} \\ \sqrt{1-\rho}e^{2\pi ki/N} & -\sqrt{\rho}e^{2\pi ki/N} \end{pmatrix}. \tag{10}$$
The solution of (9) is,
$$|\widetilde{\psi}_N(k,t)\rangle = \widetilde{U}(k)^t|\widetilde{\psi}_N(k,0)\rangle, \tag{11}$$
where $|\widetilde{\psi}_N(k,0)\rangle$ is the Fourier transformation of the initial state. To evaluate the powers of the propagator $\widetilde{U}(k)$, it is convenient to diagonalize $\widetilde{U}(k)$ using its eigenvalues and eigenstates. Using $E_n(k)$ and $|q_n(k)\rangle$ to represent the $n$th eigenvalue and orthonormalized eigenvector of the propagator ($\widetilde{U}(k)|q_n(k)\rangle = E_n(k)|q_n(k)\rangle$, $n=1,2$), Eq. (11) can be written as,
$$|\widetilde{\psi}_N(k,t)\rangle = \sum_{i=1}^{2} E_i^t(k)\langle q_i(k)|\widetilde{\psi}_N(k,0)\rangle|q_i(k)\rangle. \tag{12}$$

In the above Equation, Fourier transformation of the initial states is given by $|\widetilde{\psi}_N(k,0)\rangle = \frac{1}{\sqrt{N}}e^{2\pi ikx_0/N}|C_0\rangle$, where the initial coin state $|C_0\rangle \equiv \sqrt{a}|e_1\rangle + \sqrt{1-a}e^{i\phi}|e_2\rangle$. By performing the inverse Fourier transformation, we obtain the particle state in position representation as follows,
$$\begin{aligned}|\psi_N(x,t)\rangle &= \frac{1}{\sqrt{N}}\sum_{k=1}^{N} e^{-2\pi ikx/N}|\widetilde{\psi}_N(k,t)\rangle \\ &= \frac{1}{N}\sum_{k=1}^{N} e^{-2\pi ik(x-x_0)/N}\sum_{j=1}^{2} E_j^t(k)\langle q_j(k)|C_0\rangle|q_j(k)\rangle.\end{aligned} \tag{13}$$
Finally, we get the probability distribution,
$$\begin{aligned}P(x,t) &= |\langle e_1|\psi_N(x,t)\rangle|^2 + |\langle e_2|\psi_N(x,t)\rangle|^2 \\ &= \frac{1}{N^2}\sum_{m=1}^{2}|\sum_{k=1}^{N} e^{-2\pi ik(x-x_0)/N} \\ &\quad \times \sum_{j=1}^{2} E_j^t(k)\langle e_m|q_j(k)\rangle\langle q_j(k)|C_0\rangle|^2.\end{aligned} \tag{14}$$

Substituting eigenvalues $E_j(k)$ and eigenvectors $|q_j(k)\rangle$ (See Eq. (B1) in Appendix B) into the above equation, we obtain the probability distribution for DTQWs on 1D lattice. We have performed numerical implementations which confirm the prediction of Eq. (14). It is evident that the probability distribution depends on the initial coin states $|C_0\rangle$, eigenvalues and eigenstates of $\widetilde{U}(k)$. In the following, we will use the above equation to report the time evolution of the probability distribution for different initial states and coin parameters.

### B. Time evolution

In this section, we explore the probability distribution according to Eq. (14). Specifically, we consider the following initial coin states (a) $|C_0\rangle = |e_1\rangle$, (b) $|C_0\rangle = \frac{1}{\sqrt{2}}(|e_1\rangle \pm |e_2\rangle)$ and (c) $|C_0\rangle = \frac{1}{\sqrt{2}}(|e_1\rangle \pm i|e_2\rangle)$ with $\rho = 3/4$, $\rho = 1/2$ and $\rho = 1/4$.

Figure 1 shows the probability distribution $P(x,t)$ at $t=20$ on 1D lattice of size $N=40$. We note that the initial coin states $|C_0\rangle$ give strong influence to evolution of $P(x,t)$. For the initial states (a) and (b), $P(x,t)$ is asymmetric at the initial states. On the contrary, for initial state (c), $P(x,t)$ is symmetric and displays the same distribution for different initial phase $\phi = \pm\pi/2$. We also note from the figure that, the coin parameter $\rho$ does not bias the walk; whether $P(x,t)$ symmetric or asymmetric is totally determined by the initial coin state $|C_0\rangle$.

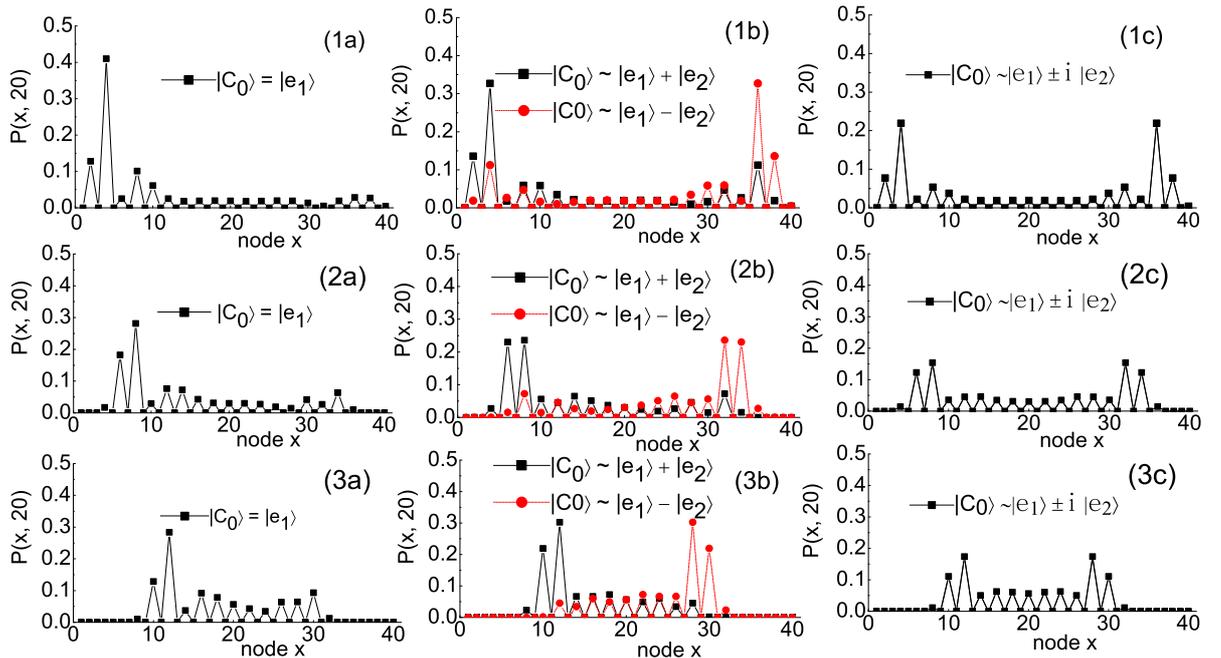

FIG. 1: (Color online) Probability distribution of DTQWs on 1D lattice of size $N = 40$ at $t = 20$. The three rows are for $\rho = 3/4$ (row 1), $\rho = 1/2$ (row 2) and $\rho = 1/4$ (row 3) while the three columns correspond to the initial state $|C_0\rangle = |e_1\rangle$ (column (a)), $|C_0\rangle = \frac{1}{\sqrt{2}}(|e_1\rangle \pm |e_2\rangle)$ (column (b)) and $|C_0\rangle = \frac{1}{\sqrt{2}}(|e_1\rangle \pm i|e_2\rangle)$ (column (c)). The initial node is at $x_0 = N/2 = 20$.

Another interesting observation is that the velocities of the two counterpropagating peaks is different for different values of $\rho$. We find that the peaks spread faster for large values of $\rho$ (Compare the row (1)-(3) in Fig. (1)). This result is consistent with the result in Ref. [32] where the positions of the peaks vary linearly with the time steps $t$. In Ref. [32], the authors show that the peak velocity $v$ is a constant value ($v \propto \sqrt{\rho}$) and differs in sign for the two directions.

It is instructive to consider the extreme values of parameter $\rho$. If $\rho = 0$, the coin flip operation $\hat{C}$ becomes the Pauli X operation, the two states cross each other going back and forth, thereby remaining close to initial excited node. If $\rho = 1$, the coin flip operation $\hat{C}$ becomes the Pauli Z operation, the two superposition states $|e_1\rangle$ and $|e_2\rangle$ move away from each other without any diffusion and interference. These two extreme cases are not of much importance, but they define the limits of the behavior. Intermediate values of $\rho$ between these extremes show intermediate drifts and quantum interference.

We also studied the evolution of probability distribution $P(x,t)$ on different lattice size,

initial states and coin parameters. The results are analogous to the case we have shown. The probability distribution generated by DTQWs consists of two counterpropagating peaks. Between the two dominant peaks the probability decays like $t^{-1}$ while outside the peaks the decay is exponential. The probability distribution $P(x,t)$ exhibits symmetric or asymmetric characteristics depending on the initial coin states.

## C. Return probabilities

Now, we consider a particular case of the probability distribution, return probability $P(x = x_0, t)$, which means the probability of finding the walker at the initial node. In order to study the scaling behavior of $P(x = x_0, t)$, we consider return probability on infinite size of lattice. In the continuum limit $N \to \infty$, $\theta(k) = 2\pi k/N$ in the eigenvalues and eigenstates (See Eq. (B1)in Appendix B) become quasicontinuous, the return probability $P(x = x_0, t)$ in Eq. (14) can be written as an integral form,

$$\begin{aligned} P(x = x_0, t)|_{N \to \infty} &= \tfrac{1}{2\pi} \sum_{m=1}^{2} |\int_{-\pi}^{\pi} (\sum_{j=1}^{2} \langle e_m | q_j(\theta) \rangle \\ &\quad \times \langle q_j(\theta) | C_0 \rangle E_j^t(\theta)) d\theta|^2 \\ &= \tfrac{1}{2\pi} \sum_{m=1}^{2} |\int_{-\pi}^{\pi} \langle e_m | q_1(\theta) \rangle \langle q_1(\theta) | C_0 \rangle e^{-it\omega(\theta)} d\theta \\ &\quad + \int_{-\pi}^{\pi} \langle e_m | q_2(\theta) \rangle \langle q_2(\theta) | C_0 \rangle (-1)^t e^{it\omega(\theta)} d\theta|^2 \end{aligned} \quad (15)$$

where $E_1 = e^{-i\omega(\theta)}$, $E_2 = -e^{i\omega(\theta)}$ and $\omega(\theta) = arcsin(\sqrt{\rho} \sin \theta)$ are applied in the above equation.

In Appendix B, we use the stationary phase approximation (SPA) (See Appendix A) to calculate the above integral. We find that the integral is finally simplified as,

$$P(x = x_0, t)|_{N \to \infty} = \begin{cases} \frac{2\sqrt{1/\rho - 1}}{\pi t}, & \text{if } t \in Even, \\ 0 & \text{if } t \in Odd. \end{cases} \quad (16)$$

Equation (16) indicates that the return probability show scaling behavior $P(x = x_0, t)|_{N \to \infty} \sim t^{-1}$. We note that the return probability does not depend on the parameters ($a$ and $\phi$ in $|C_0\rangle$, see Eq. (6)) of the initial states. Particularly, when $\rho = 1/2$, we obtain the return probability for Hadamard walks: $P(x_0, t)|_{N \to \infty} = 2/(\pi t)$. The parameter $\rho$ only affects the coefficient of the scaling $t^{-1}$. The scaling behavior of $P(x_0, t)|_{N \to \infty}$ is analogous to the return probability of continuous-time quantum walks. For continuous-time quantum walks on 1D lattice, the return probability is given by $\pi(t) = |J_0(2t)|^2 \approx \sin^2(2t + \pi/4)/(\pi t)$, where $J_n(x)$ is the Bessel function of the first kind [33, 34]. Thus, both the return probability of the discrete-time and continuous-time quantum walks show the same scaling behavior $t^{-1}$.

In order to test the prediction of Eq. (16), Fig. 2 shows the return probability $P(x = x_0, t)$ on a 1D lattice of size $N = 200$ with $\rho = 1/4$, $\rho = 1/2$ and $\rho = 3/4$ for the first 100 steps. In our calculation, we fix the value of $\rho$ and try to change the initial states of the walks, and find that the return probabilities are exact the same. This confirms our conclusion that the return probability is independent on the initial coin states. We also show the theoretical predictions of Eq. (16) in Fig. 2, which are in good agreement with the numerical results.

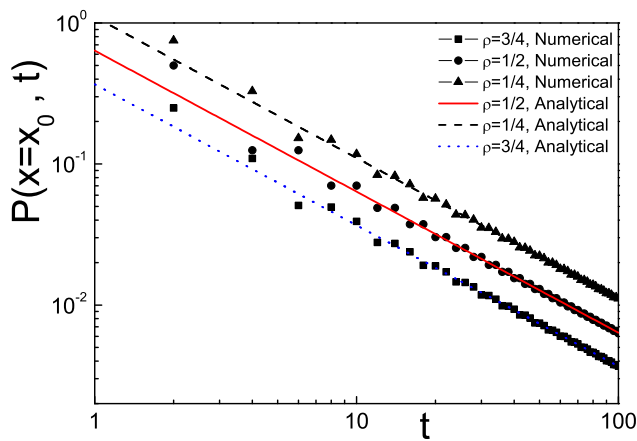

FIG. 2: (Color online) Return probability $P(x = x_0, t)$ on 1D lattice of size $N = 200$ with $\rho = 1/4$ (triangles), $\rho = 1/2$ (dots) and $\rho = 3/4$ (squares) in the first 100 steps. The lines show the predictions of Eq. (16) for $\rho = 1/4$ (dashed line), $\rho = 1/2$ (solid line) and $\rho = 3/4$ (dotted line), respectively. Since $P(x = x_0, t)$ equals to 0 at odd-numbered step $t$, we only plot $P(x = x_0, t)$ at even-numbered step $t$ in the figure.

### D. Long-time limiting probabilities

In this section, we consider long time averages of the probability distribution. Generally, the time-averaged distribution $\bar{P}(x,T) \equiv \frac{1}{T}\sum_{t=1}^{T} P(x,t)$ converges to a constant value as $T \to \infty$. This value is defined as the long-time limiting probability,

$$\chi(x) = \lim_{T \to \infty} \bar{P}(x,T). \tag{17}$$

Substituting the eigenvalues $E_1 = e^{-i\omega(\theta_k)}$ and $E_2 = -e^{i\omega(\theta_k)}$ into Eq. (14), we obtain,

$$\begin{aligned}
P(x,t) &= \frac{1}{N^2} \sum_{m=1}^{2} |\sum_{k=1}^{N} e^{-2\pi i k(x-x_0)/N} \\
&\quad \times (e^{-i\omega(k)t}\langle e_m|q_1(k)\rangle\langle q_1(k)|C_0\rangle \\
&\quad + (-1)^t e^{i\omega(k)t}\langle e_m|q_2(k)\rangle\langle q_2(k)|C_0\rangle)|^2 \\
&= \frac{1}{N^2} \sum_{m=1}^{2} \sum_{k=1,k'=1}^{N} e^{-2\pi i(k-k')(x-x_0)/N} \\
&\quad \times [e^{-i\omega(k)t}\langle e_m|q_1(k)\rangle\langle q_1(k)|C_0\rangle \\
&\quad + (-1)^t e^{i\omega(k)t}\langle e_m|q_2(k)\rangle\langle q_2(k)|C_0\rangle] \\
&\quad \times [e^{i\omega(k')t}\langle C_0|q_1(k')\rangle\langle q_1(k')|e_m\rangle \\
&\quad + (-1)^t e^{-i\omega(k')t}\langle C_0|q_2(k')\rangle\langle q_2(k')|e_m\rangle].
\end{aligned} \tag{18}$$

Our goal is to calculate the long time averages of the probability distribution. Only two terms of the product in Eq. (19) contributes if $\omega(k) = \omega(k')$, since $\lim_{T \to \infty} \frac{1}{T}\sum_{t=1}^{T} e^{\pm i(\omega(k)-\omega(k'))t} = \delta_{\omega(k),\omega(k')}$ and $\lim_{T \to \infty} \frac{1}{T}\sum_{t=1}^{T}(-1)^t e^{\pm i(\omega(k)+\omega(k'))t} = 0$. Thus, the long-time limiting probability $\chi(x)$ can be simplified as,

$$\begin{aligned}
\chi(x) &= \frac{1}{N^2} \sum_{m=1}^{2} \sum_{k=1,k'=1}^{N} \delta_{\omega(k),\omega(k')} e^{-2\pi i(k-k')(x-x_0)/N} \\
&\quad \times [\langle e_m|q_1(k)\rangle\langle q_1(k)|C_0\rangle\langle C_0|q_1(k')\rangle\langle q_1(k')|e_m\rangle \\
&\quad + \langle e_m|q_2(k)\rangle\langle q_2(k)|C_0\rangle\langle C_0|q_2(k')\rangle\langle q_2(k')|e_m\rangle]
\end{aligned} \tag{19}$$

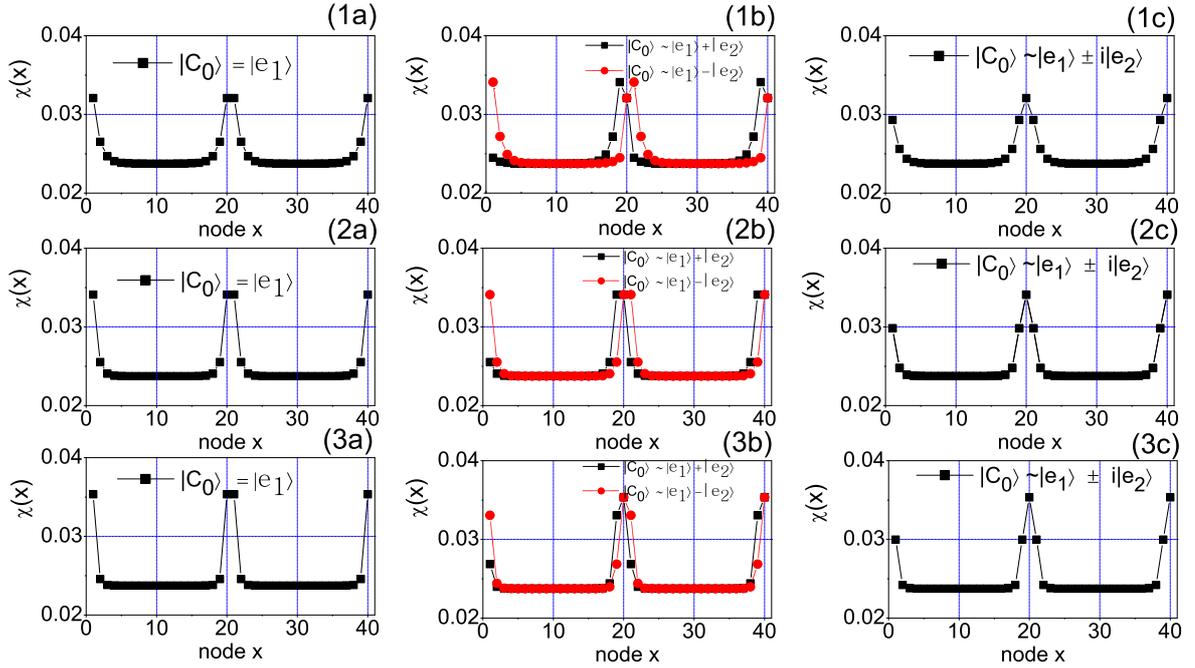

FIG. 3: (Color online) Long-time limiting probability $\chi(x)$ on 1D lattice of size $N = 40$ for $\rho = 3/4$ (row 1), $\rho = 1/2$ (row 2) and $\rho = 1/4$ (row 3) with initial state $|C_0\rangle = |e_1\rangle$ (column (a)), $|C_0\rangle = \frac{1}{\sqrt{2}}(|e_1\rangle \pm |e_2\rangle)$ (column (b)) and $|C_0\rangle = \frac{1}{\sqrt{2}}(|e_1\rangle \pm i|e_2\rangle)$ (column (c)). The walk starts at the initial node $x_0 = N/2 = 20$.

$\chi(x)$ is dependent on the initial coin states $|C_0\rangle$ and coin parameter $\rho$ (note the eigenstates $|q(k)\rangle$ depends on $\rho$). Here, we report the long-time limiting probabilities for various initial states $|C_0\rangle$ and coin parameter $\rho$.

We consider the following initial coin states (a) $|C_0\rangle = |e_1\rangle$, (b) $|C_0\rangle = \frac{1}{\sqrt{2}}(|e_1\rangle \pm |e_2\rangle)$ and (c) $|C_0\rangle = \frac{1}{\sqrt{2}}(|e_1\rangle \pm i|e_2\rangle)$ with $\rho = 3/4$, $\rho = 1/2$ and $\rho = 1/4$. Fig. 3 shows the long-time limiting probabilities $\chi(x)$ on 1D lattice of size $N = 40$. For initial state (c), $\chi(x)$ is a symmetric distribution, i.e., $\chi(x) = \chi(x')$ if $x - x_0 = x_0 - x'$. $\chi(x)$ also displays localization on the nodes nearing the initial node $x_0$ and the opposite node $\hat{x}_0 \equiv x_0 + N/2 \pmod{N}$. For the same initial state, it seems that small values of $\rho$ lead to a strong localization than large values of $\rho$ (compare different rows in the same column). For initial state (a), $\chi(x)$ has large values at the initial node $x_0$ (or opposite node $\hat{x}_0$) and its next node $x_0 + 1$ (or $\hat{x}_0 + 1$). These high probabilities are exactly equal, i.e., $\chi(x_0) = \chi(x_0 + 1) = \chi(\hat{x}_0) = \chi(\hat{x}_0 + 1)$. A similar phenomena is also observed for the initial state (b) (See column (b) in Fig 3). $\chi(x)$ has large values at the $x_0$'s (or $\hat{x}_0$'s) previous or next nodes, depending on the sign of the initial state. Here, $\chi(x_0) = \chi(x_0 + 1) = \chi(\hat{x}_0) = \chi(\hat{x}_0 + 1)$ holds only when $\rho = 1/2$ (See Fig. (3(2b))). The probabilities are equal at nodes $x$ and $x + N/2 \pmod{N}$ for all the initial states and values of $\rho$. This behavior is a natural consequence of the periodic boundary conditions of the 1D lattice. The limiting probability for initial state (c) is symmetric at the origin node $x_0$, i.e., $\chi(x - x_0) = \chi(x_0 - x)$. However, this is not true for conditions (b) and

(c). This feature is related to the initial coin states. Generally, symmetric initial state leads to unbiased limiting probability distribution while asymmetric coin state produces biased quantum walks.

In the above analysis, we have studied the limiting probabilities on lattice of size $N = 40$. For odd values of $N$, we find that $\chi(x)$ is a uniform distribution for all the initial states and values of $\rho > 0$. For even values of $N$, we find that the distribution pattern depends on the parity of $N/2$. If $N/2$ is an even number, $\chi(x)$ has peaks at the nodes nearing origin node and the opposite node. However, if $N/2$ is an odd number, $\chi(x)$ has a peak nearing the origin node but a minimum nearing the opposite node. The minimum in odd $N/2$ is virtually the mirror of the peak nearing the origin node $x_0$ (See Fig. 1 in Ref. [22]). Nevertheless, the situation is different for the extreme case $\rho = 1$ and $\rho = 0$. If $\rho = 1$, two superposition states move away from each other without any diffusion and interference, the limiting probability is a uniform distribution for all the initial states and values of $N$. If $\rho = 0$, the limiting probability $\chi(x)$ is total determined by the initial states $|C_0\rangle \equiv (\sqrt{a}|e_1\rangle + \sqrt{1-a}e^{i\phi}|e_2\rangle)$. More specifically, the limiting probability $\chi(x)$ for $\rho = 0$ is summarized as,

$$\chi(x) = \begin{cases} a/2, & \text{if } x = x_0 + 1, \\ 1/2, & \text{if } x = x_0, \\ (1-a)/2, & \text{if } x = x_0 - 1, \\ 0, & Otherwise \end{cases} \quad (20)$$

It is worth mentioning that the limiting probability distributions between the continuous-time and discrete-time quantum walks are different. For continuous-time quantum walks (CTQWs), the limiting probability only depends on the parity of $N$ (See Eq.(21) and (22) in Ref. [34]). On the contrary, the limiting probability for discrete-time quantum walks depends on more ingredients. This is because the coin degrees of freedom of DTQWs offer a wider range of controls over the evolution of the walk than the continuous-time quantum walk.

## IV. MIXING TIME

As mentioned in the previous section, the time averaged probability distribution $\bar{P}(x,T)$ converges to the limiting probability $\chi(x)$ as $T \to \infty$. Here, we study this issue using the concept of mixing time. Mixing time represents the rate at which the average probability distribution $\bar{P}(x,T)$ approaches its asymptotic distribution $\chi(x)$. The average mixing time is defined as follows,

$$M_\epsilon = min\{ \tau \mid \forall T > \tau, \sum_x |\bar{P}(x,T) - \chi(x)| < \epsilon\}. \quad (21)$$

Fig. 4 (a) shows the time dependence of the variation distance $V(T) \equiv \sum_x |\bar{P}(x,T) - \chi(x)|$ on 1D lattices of size $N = 10$, $N = 20$ and $N = 100$. For long times, the variation distance $V(T)$ oscillate frequently and decays approximately as $1/T$. For odd-numbered size of $N$, the probability mixes to the uniform distribution, we also find a similar behavior of $V(T)$ (See $V(T)$ vs $T$ for $N = 11$, $N = 21$ and $N = 101$ in Fig. 4 (b)).

Fig. 5 shows the dependence of the average mixing time $M_\epsilon$ on the lattice size $N$ with different values of threshold $\epsilon$, for the initial state $|C_0\rangle = \frac{1}{\sqrt{2}}(|e_1\rangle \pm i|e_2\rangle)$. For sufficiently large $\epsilon$, the average mixing time is a linear function of $N$. However, for small values of $\epsilon$, $M_\epsilon$ shows wild fluctuation around the linear behavior $M_\epsilon \propto N$.

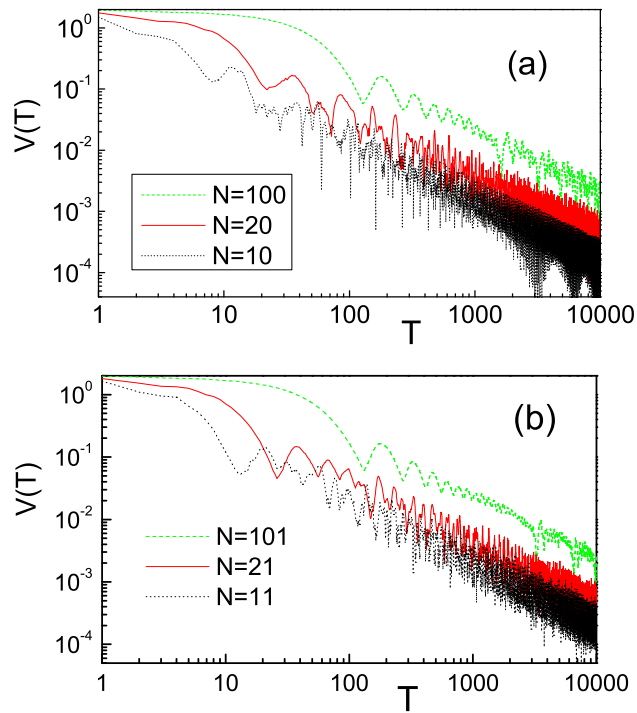

FIG. 4: (Color online) Variation distance $V(T)$ as a function of time $T$ for 1D lattices of size $N = 10$ (dotted curve), $N = 20$ (solid curve) and $N = 100$ (dashed curve). The results are for Hadamard quantum walks ($\rho = 1/2$) with symmetric initial state $|C_0\rangle = \frac{1}{\sqrt{2}}(|e_1\rangle \pm i|e_2\rangle)$.

We also try to compare the average mixing time $M_\epsilon$ for different values of $\rho$ and initial states $|C_0\rangle$. We consider quantum walks for the initial states (a), (b), (c) with $\rho = 3/4$, $\rho = 1/2$ and $\rho = 1/4$. We find that quantum walk for initial states (c) with $\rho = 3/4$ has a smaller mixing time $M_\epsilon$ than the other cases considered here. This may suggest that quantum walks with symmetric initial states and large values of $\rho$ mix to the limiting probability distribution fast. We hope this conclusion can be used in constructing efficient quantum algorithms.

## V. NEW QUANTUM WALK ON 1D LATTICE

In this section, we introduce another kind of quantum walk on 1D lattice. The quantum walk is defined on an infinite or even-numbered size of lattice. The walk starts at node $x_0$ with initial coin state $C_0|\rangle = a_0|e_1\rangle + b_0|e_2\rangle$, we endow "direction" to the edges in the graph. We label each edge a direction ($|e_1\rangle$ or $|e_2\rangle$), so that the edges incident on every node can be labeled as two different directions and every edge between two nodes has the same label at either end. Only 1D lattices of even-numbered size satisfy this condition. We illustrate this kind of labeling in Fig. 6. The walk is evolved into the superposition of the coin space

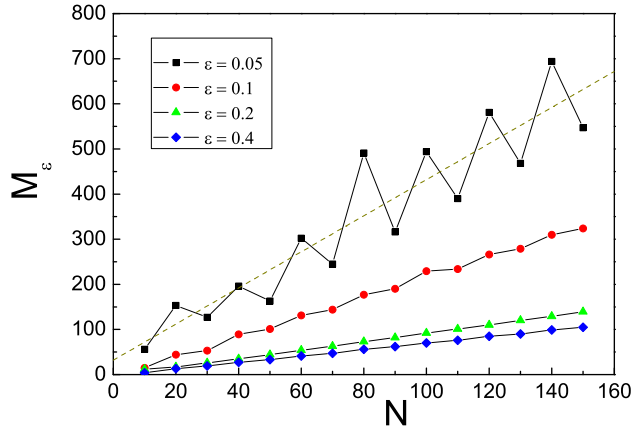

FIG. 5: (Color online) Dependence of mixing time $M_\epsilon$ on the lattice size $N$ with $\epsilon = 0.05$, $\epsilon = 0.1$, $\epsilon = 0.2$ and $\epsilon = 0.4$. The results are obtained using $\rho = 1/2$ and symmetric initial state $|C_0\rangle = \frac{1}{\sqrt{2}}(|e_1\rangle \pm i|e_2\rangle)$.

by applying the coin flip operation, then the shift operation $\hat{S}'$ moves the walker according to
$$\begin{aligned}\hat{S}' &= [\sum_{x \in G_1}(|x-1\rangle\langle x| \otimes |e_1\rangle\langle e_1| + |x+1\rangle\langle x| \otimes |e_2\rangle\langle e_2|)] \\ &+ [\sum_{x \in G_2}(|x+1\rangle\langle x| \otimes |e_1\rangle\langle e_1| + |x-1\rangle\langle x| \otimes |e_2\rangle\langle e_2|)] \\ &\equiv \hat{S}_1 + \hat{S}_2,\end{aligned} \quad (22)$$

where we use $\hat{S}_1$ and $\hat{S}_2$ to denote the two terms in the above equation. We separate the total shift operation $\hat{S}$ into two different flip operators, $\hat{S}_1$ and $\hat{S}_2$, which are applied to two different node group $G_1 = \{x|..., x = x_0 - 4, x = x_0 - 2, x = x_0, x = x_0 + 2, x = x_0 + 4, ...\}$ and $G_2 = \{x|..., x = x_0 - 3, x = x_0 - 1, x = x_0 + 1, x = x_0 + 3, ...\}$, respectively. We implement the above process iteratively to realize a large number of steps of the quantum walk. The peculiarity of this walk distinguished from the traditional quantum walk is the conditional shift operation $\hat{S}$. In the traditional quantum walk, the shift operation $\hat{S}$ moves the walker to the same side of the node, regardless of the position of the walker. However, in our defined quantum walk, the shift operation $\hat{S}$ moves the walker toward different sides, depending on the position of walker.

A natural question is "what's the relationship between the two kinds of quantum walks?". To answer this question, we consider the two quantum walks on the same 1D lattice. For the sake of simplicity, we compare the wave function of the two quantum walks with different initial states and coin flip matrixes. Concretely, we consider the traditional quantum walk with initial coin state $|C_0\rangle = a_0|e_1\rangle + b_0|e_2\rangle$ and shift matrix $\hat{C}$ in Eq. (5), and our defined quantum walk with initial coin state $|C_0'\rangle = b_0|e_1\rangle + a_0|e_2\rangle$ and shift matrix $\hat{C}' = \begin{pmatrix} \sqrt{1-\rho} & \sqrt{\rho} \\ \sqrt{\rho} & -\sqrt{1-\rho} \end{pmatrix}$. After $t$ steps, suppose the state of the traditional quantum walk at node $x$ is $|\psi(x,t)\rangle = (A_{x,t}|e_1\rangle + B_{x,t}|e_2\rangle)|x\rangle$, then the state of our defined quantum walk

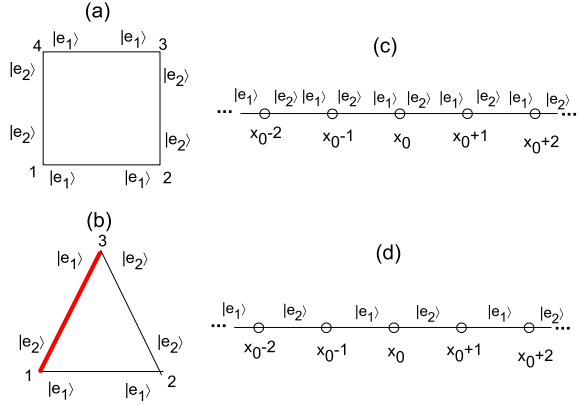

FIG. 6: (Color online) Illustration of the edge labeling on lattice of size $N = 4$ (a) and $N = 3$ (b). In (a), the edges $1 \leftrightarrow 2$, $3 \leftrightarrow 4$ are labeled as direction $|e_1\rangle$, edges $2 \leftrightarrow 3$, $1 \leftrightarrow 4$ are labeled as direction $|e_2\rangle$. In (b), the edges $1 \leftrightarrow 2$, $2 \leftrightarrow 3$ are labeled as direction $|e_1\rangle$ and $|e_2\rangle$, but edge $1 \leftrightarrow 3$ can't be labeled as a single direction. (c) The labeling of coin direction for the traditional quantum walk, where the left side of each node is labeled as $|e_1\rangle$ and right side labeled as $|e_2\rangle$. (d) The labeling of our defined quantum walk. $|e_1\rangle$ and $|e_2\rangle$ are labeled to each edge and two nodes of the edge has the same label.

$|\psi'(x,t)\rangle$ is given by,

$$|\psi'(x,t)\rangle = \begin{cases} (B_{x,t}|e_1\rangle + A_{x,t}|e_2\rangle)|x\rangle, x \in G_1 \text{ if } t \in Even, \\ (A_{x,t}|e_1\rangle - B_{x,t}|e_2\rangle)|x\rangle, x \in G_2 \text{ if } t \in Odd. \end{cases} \quad (23)$$

The above conclusion can be proved using the method of mathematical induction. For $t = 0$ and $t = 1$, it is easy to see that the wave functions satisfy the above equation. Now suppose for $t = T_0$ ($T_0 > 1$) the conclusion also holds, then we obtain the wave function $|\psi(x, T_0 + 1)\rangle$ at $t = T_0 + 1$ according to the iterative relations,

$$\begin{aligned} |\psi(x, T_0 + 1)\rangle &= (\sqrt{\rho}A_{x+1,T_0} + \sqrt{1-\rho}B_{x+1,T_0})|e_1\rangle|x\rangle \\ &+ (\sqrt{1-\rho}A_{x-1,T_0} - \sqrt{\rho}B_{x-1,T_0})|e_2\rangle|x\rangle \\ &\equiv A_{x,T_0+1}|e_1\rangle|x\rangle + B_{x,T_0+1}|e_2\rangle|x\rangle \end{aligned} \quad (24)$$

where we use $A_{x,T_0+1}$ and $B_{x,T_0+1}$ to represent the first two terms. Applying $\hat{S}'(\hat{I} \otimes \hat{C}')$ to $|\psi'(x, T_0)\rangle$ and using the iterative relations, we obtain the wave function $|\psi'(x, T_0 + 1)\rangle$ at $t = T_0 + 1$

$$|\psi'(x, T_0 + 1)\rangle = \begin{cases} (\sqrt{\rho}A_{x+1,T_0} + \sqrt{1-\rho}B_{x+1,T_0})|e_1\rangle|x\rangle + \\ (-\sqrt{1-\rho}A_{x-1,T_0} + \sqrt{\rho}B_{x-1,T_0})|e_2\rangle|x\rangle, x \in G_2 \text{ if } T_0 \in Even, \\ (\sqrt{1-\rho}A_{x-1,T_0} - \sqrt{\rho}B_{x-1,T_0})|e_1\rangle|x\rangle + \\ (\sqrt{\rho}A_{x+1,T_0} + \sqrt{1-\rho}B_{x+1,T_0})|e_1\rangle|x\rangle. \ x \in G_1 \quad \text{if } T_0 \in Odd. \end{cases} \quad (25)$$

Comparing Eqs. (24) and (26), we have

$$|\psi'(x, T_0 + 1)\rangle = \begin{cases} A_{x,T_0+1}|e_1\rangle|x\rangle - B_{x,T_0+1}|e_2\rangle|x\rangle & x \in G_2, \text{ if } T_0 \in Even, \\ B_{x,T_0+1}|e_1\rangle|x\rangle + A_{x,T_0+1}|e_2\rangle|x\rangle, & x \in G_1, \text{ if } T_0 \in Odd. \end{cases} \quad (26)$$

Therefore, our conclusion is also true for $t = T_0 + 1$. According to the mathematical induction, the relation holds for all the time steps. This completes our proof.

According to the deduction, the two types of quantum walks have the same probability distribution. This indicates that our defined quantum walk is equivalent to the traditional quantum walk with a symmetrical initial state and coin parameter $\rho$. We have performed numerical implementations to realize the quantum walks defined here, and the results support our findings.

## VI. CONCLUSIONS

In summary, we have studied discrete-time quantum walks on one-dimensional lattices. We show that the evolution of the quantum dynamics depends on the initial states and coin parameters. For infinite size of lattice, we derive an explicit expression for the return probability, which shows scaling behavior $P(x = x_0, t) \sim t^{-1}$ and does not depends on the initial states of the walk. In the long-time limit, the probability distribution shows various patterns, depending on the initial states, coin parameters and the lattice size. The average mixing time $M_\epsilon$ closes to the limiting probability in linear $N$ for large values of thresholds $\epsilon$. Finally, we define another kind of quantum walk on infinite or even-numbered size of lattices, and find that the walk is equivalent to the traditional quantum walk with symmetric initial state and coin parameter.

### Acknowledgments

This work is supported by National Natural Science Foundation of China under projects 10975057 and the new Teacher Foundation of Suzhou University.

### APPENDIX A: THE STATIONARY PHASE APPROXIMATION (SPA)

Stationary phase approximation (SPA) is an approach for solving integrals analytically by evaluating the integrands in regions where they contribute the most [35–37]. This method is specifically directed to evaluating oscillatory integrands, where the phase function of the integrand is multiplied by a relatively high value. Suppose we want to evaluate the behavior of function $I(\lambda)$ for large $\lambda$,

$$I(\lambda) = \frac{1}{2\pi} \int g(x) e^{-\lambda f(x)} dx. \tag{A1}$$

The SPA asserts that the main contribution to this integral comes from those points where $f(x)$ is stationary $[df(x)/dx \equiv f'(x) \equiv 0]$. If there is only one point $x_0$ for which $f'(x_0) = 0$ and $d^2 f(x)/dx^2 |_{x_0} \equiv f''(x_0) \neq 0$, the integral is approximated asymptotically by,

$$I(\lambda) \approx \frac{1}{\sqrt{2\pi \lambda f''(x_0)}} g(x_0) e^{-\lambda f(x_0)}. \tag{A2}$$

If there are more than one stationary points satisfy $[df(x)/dx \equiv f'(x) \equiv 0]$, then the integral $I(\lambda)$ is approximately given by the sum of the contributions [each being of the form given in Eq. (A2)] of all the stationary points [36].

# APPENDIX B: CALCULATION OF THE RETURN PROBABILITY USING SPA

The eigenvalues and eigenstates of $\widetilde{U}(k)$ can be written as,

$$\begin{aligned}
E_1 &= e^{-i\omega(k)}, \quad E_2 = -e^{i\omega(k)}; \\
|u_1\rangle &= [2 - 2\sqrt{\rho}\cos(\omega(k) - \theta(k))] \\
&\quad \times \{\sqrt{1-\rho}|e_1\rangle + [-\sqrt{\rho} + e^{-i(\omega(k)-\theta(k))}]|e_2\rangle\}, \\
|u_2\rangle &= [2 + 2\sqrt{\rho}\cos(\theta(k) + \omega(k))] \\
&\quad \times \{\sqrt{1-\rho}|e_1\rangle + [-\sqrt{\rho} - e^{i(\omega(k)+\theta(k))}]|e_2\rangle\}, \\
|q_1\rangle &= \frac{|u_1\rangle}{\langle u_1|u_1\rangle}, \quad |q_2\rangle = \frac{|u_2\rangle}{\langle u_2|u_2\rangle};
\end{aligned} \quad (B1)$$

where $\theta(k) = 2\pi k/N$ and $\sin\omega(k) = \sqrt{\rho}\sin\theta(k)$.

In the continuum limit $N \to \infty$, the values of $\theta_k = 2\pi k/N$ are quasicontinuous, then the return probability $P(x = x_0, t)$ can be written as the integral form in Eq. (15). Now we apply SPA to calculate this integral. In Eq. (15), $P(x = x_0, t)|_{N\to\infty}$ can be written as,

$$P(x = x_0, t)|_{N\to\infty} = I_1 + I_2, \quad (B2)$$

where

$$\begin{aligned}
I_1 &= \frac{1}{2\pi}|\int_{-\pi}^{\pi}\langle e_1|q_1(\theta)\rangle\langle q_1(\theta)|C_0\rangle e^{-it\omega(\theta)}d\theta \\
&\quad + \int_{-\pi}^{\pi}\langle e_1|q_2(\theta)\rangle\langle q_2(\theta)|C_0\rangle(-1)^t e^{it\omega(\theta)}d\theta|^2, \\
I_2 &= \frac{1}{2\pi}|\int_{-\pi}^{\pi}\langle e_2|q_1(\theta)\rangle\langle q_1(\theta)|C_0\rangle e^{-it\omega(\theta)}d\theta \\
&\quad + \int_{-\pi}^{\pi}\langle e_2|q_2(\theta)\rangle\langle q_2(\theta)|C_0\rangle(-1)^t e^{it\omega(\theta)}d\theta|^2.
\end{aligned} \quad (B3)$$

The stationary points of the above integrals satisfy $\omega'(\theta) = d\arcsin(\rho\sin\theta)/d\theta = \sqrt{\rho}\cos(\theta)/\sqrt{1-\rho\sin^2(\theta)} = 0$. So the contribution of each integral comes from two stationary points $\theta = \pi/2$ and $\theta = -\pi/2$. The second-order derivations at $\theta = \pm\pi/2$ yield $\omega''(\pm\pi/2) = \mp\sqrt{\rho}/\sqrt{1-\rho}$. According to the SPA and substituting $\theta = \pm\pi/2$ into Eqs. (B1)-(B3), we obtain the integral $I_1$ and $I_2$ as follows,

$$I_1 \approx |z_1 + z_2 + z_3 + z_4|^2, \quad (B4)$$

where

$$\begin{aligned}
z_1 &= \frac{(1/\rho - 1)^{1/4}}{2\sqrt{-2\pi it}}(\sqrt{a} + i\sqrt{1-a}e^{i\phi})e^{-it\omega_0}, \\
z_2 &= -i\frac{(1/\rho - 1)^{1/4}}{2\sqrt{-2\pi it}}(\sqrt{a} - i\sqrt{1-a}e^{i\phi})e^{it\omega_0}, \\
z_3 &= (-1)^t z_2, \quad z_4 = (-1)^t z_1.
\end{aligned} \quad (B5)$$

where $\omega_0 = \arcsin\sqrt{\rho}$. And

$$I_2 \approx |f_1 + f_2 + f_3 + f_4|^2, \quad (B6)$$

where

$$f_1 = iz_1, f_2 = -iz_2, f_3 = (-1)^t f_2, f_4 = (-1)^t f_1. \tag{B7}$$

If $t$ is odd, $z_1 + z_4 = z_2 + z_3 = 0$ and $f_1 + f_4 = f_2 + f_3 = 0$, the integral equals to 0. If $t$ is even, $z_1 = z_4$, $z_2 = z_3$, $f_1 = f_4$, $f_2 = f_3$, the integral is simplified as,

$$\begin{aligned} I &= I_1 + I_2 = |2(z_1 + z_2)|^2 + |2(f_1 + f_2)|^2 \\ &= \frac{\sqrt{1/\rho - 1}}{2\pi t} \times \\ & [|(\sqrt{a} + i\sqrt{1-a}e^{i\phi})e^{-it\omega_0} + (-i\sqrt{a} - \sqrt{1-a}e^{i\phi})e^{it\omega_0}|^2 \\ & + |(i\sqrt{a} - \sqrt{1-a}e^{i\phi})e^{-it\omega_0} + (-\sqrt{a} + i\sqrt{1-a}e^{i\phi})e^{it\omega_0}|^2] \\ &= \frac{\sqrt{1/\rho - 1}}{2\pi t} \times 4 \\ &= \frac{2\sqrt{1/\rho - 1}}{\pi t} \end{aligned} \tag{B8}$$

Therefore, we obtain the return probability

$$P(x = x_0, t)|_{N \to \infty} = \begin{cases} \frac{2\sqrt{1/\rho - 1}}{\pi t}, & \text{if } t \in Even, \\ 0 & \text{if } t \in Odd. \end{cases} \tag{B9}$$

---